\documentclass[sigconf,screen]{acmart}

\acmBooktitle{Proceedings of the 32nd ACM Symposium on the Foundations of Software Engineering (FSE '24), November 15--19, 2024, Porto de Galinhas, Brazil}

\AtBeginDocument{%
  \providecommand\BibTeX{{%
    \normalfont B\kern-0.5em{\scshape i\kern-0.25em b}\kern-0.8em\TeX}}}

\setcopyright{acmcopyright}
\copyrightyear{2024}
\acmYear{2024}
\acmDOI{XXXXXXX.XXXXXXX}

\acmPrice{15.00}
\acmISBN{978-1-4503-XXXX-X/18/06}

\usepackage{eiffel}
\usepackage{soul}
\usepackage{enumitem}
\setitemize{leftmargin=*}

\newcommand{\e}[1]{\mbox{\lstinline[basicstyle=\normalsize]|#1|}}

\newcommand{\foote}[1]{\mbox{\color[HTML]{222277}\footnotesize\bfseries\ttfamily}}

\begin{document}

    \title{Execution-free program repair}

\author{Li Huang}
\email{Li.Huang@constructor.org}
\orcid{1234-5678-9012}

\author{Bertrand Meyer}
\email{Bertrand.Meyer@inf.ethz.ch}
\orcid{0000-0002-5985-7434}

\author{Ilgiz Mustafin}
\email{Ilgiz.Mustafin@constructor.org}
\orcid{0009-0007-0476-5966}

\author{Manuel Oriol}
\email{mo@constructor.org}
\orcid{0000-0003-4069-7626}

\affiliation{%
  \institution{Constructor Institute}
  \streetaddress{Rheinweg 9}
  \city{Schaffhausen}
  \country{Switzerland }
  \postcode{8212}
}

\newcommand{\toolname}{Proof\-2\-Fix}



\begin{abstract}
[Draft of article to appear in FSE 2024, Foundations of Software Engineering, July 2024.]

Automatic program repair usually relies heavily on test cases for both bug identification and fix validation.
The issue is that writing test cases is tedious, running them takes much time, and validating a fix through tests does not guarantee its correctness. 
The novel idea in the \toolname~methodology and tool presented here is to rely instead on a program prover, without the need to run tests or to run the program at all. 
Results show that \toolname~automatically finds and fixes significant historical bugs.
\end {abstract}




\maketitle

\section{Overview}

Finding bugs is good; correcting them is better. 
An important trend in the recent evolution of software verification research is the development of methods of Automatic Program Repair (APR), which propose corrections to bugs. 
Whereas results are promising, most of the existing APR approaches are \textit{dynamic}  \cite{monperrus2018automatic, 
monperrus2018living}: they infer and validate the corrections by running the program. The present work describes a static, \textit{prover-based} approach, \toolname, which only needs the source code to find and fix bugs.

Automatic program repair typically involves four steps \cite{le2011genprog}: identify a fault (also called a ``bug''); localize the fault; generate a fix (also called ``correction'' or ``patch''); evaluate the fix.
Moving from a dynamic test-based approach to a static prover-based approach benefits all these goals but particularly two:

\begin{itemize}
    \item \textit{Fault identification}:
    A dynamic approach needs to prepare test inputs,  often many of them, and run tests until one triggers the bug. In spite of progress towards automatic techniques \cite{pacheco2007randoop, tillmann2008pex, wei2010satisfying}, test case preparation remains, in most practical cases, a considerable labor-intensive task. 
    With the static approach presented here, one simply runs a prover on the program text; bugs are detected when the prover fails to show that the program is correct.
    \footnote{Note that the limitation of the prover sometimes prevents it to assess code correctness. With modern provers this case can largely be eliminated or worked around.} 
    No need to invent test data or produce a test harness.

    \item  \textit{Fix validation}: this step is perhaps where a prover-based approach has the most substantial advantage. 
    Once we have a candidate fix, a dynamic approach must run all the tests again and check that they pass; not only is this process time-consuming,  it is also fraught with uncertainty since a positive answer is only as good as the test suite, by nature non-exhaustive. 
    In a prover-based approach the fix validation is a proof, which provides a guarantee that the fix is correct (or not).


\end{itemize}

\noindent A prover-based approach can also help \textit{Fault localization}, by pinpointing (often down to the level of a particular instruction) the precise program element that precludes verification, and \textit{Fix generation}, by avoiding biases.

Taking advantage of a modern proof environment, the \toolname~approach and tool described in the following sections produce meaningful fixes and validates them formally.

\section{A sample session} \label{example}
\begin{figure}
  \centering
  \includegraphics[width=7.5cm]{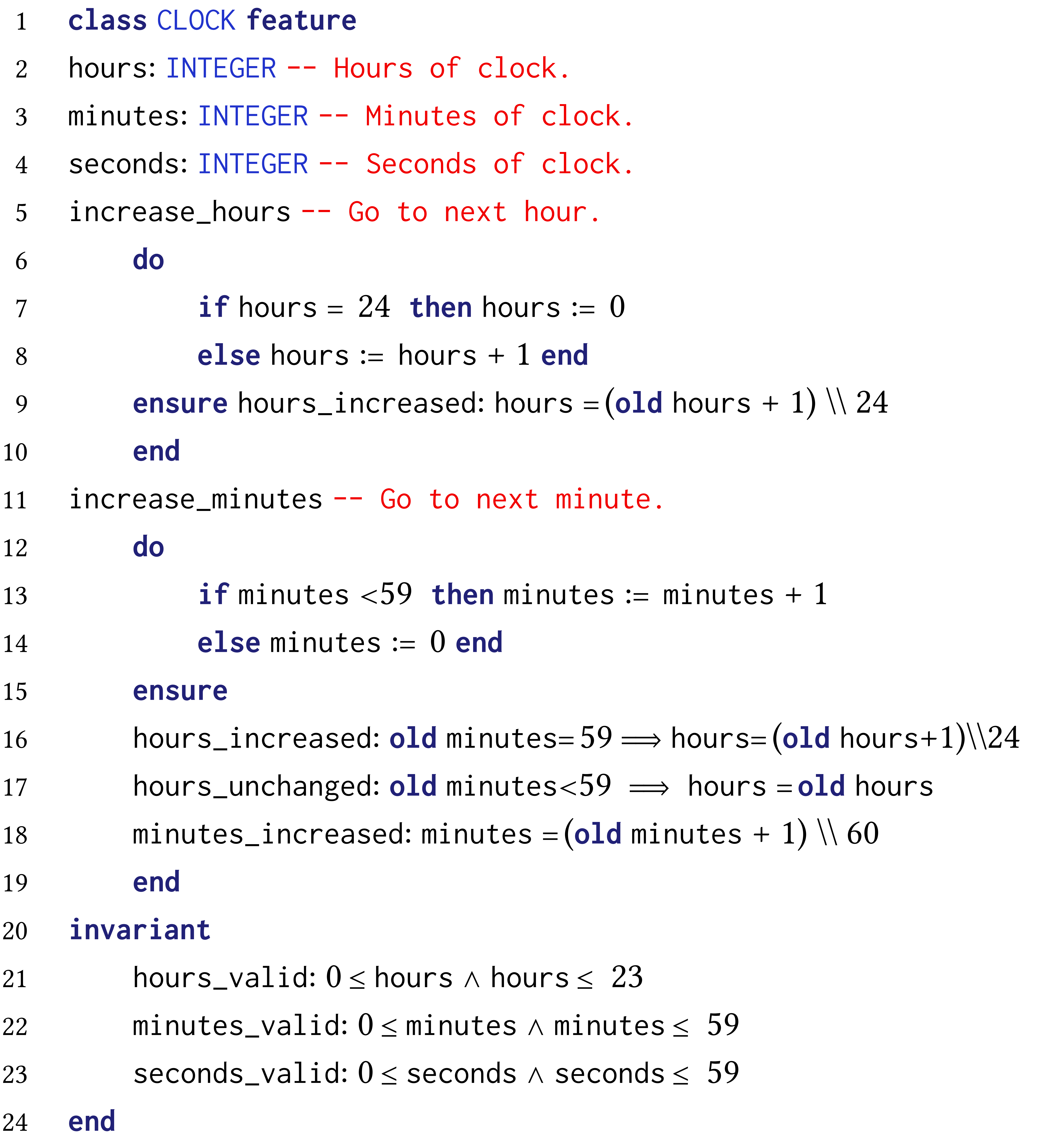}
  \vspace{-0.3cm}
\caption{A buggy version of the \e{CLOCK} class}
\label{fig: clock}
\end{figure}

To explain the main idea before going into the underlying technology, we start with a typical use of the \toolname~on a small but representative example (Fig. \ref{fig: clock}). 
The \e{CLOCK} class, in Eiffel, implements a digital clock with fields \e{seconds}, \e{minutes} and \e{hours}, and routines to increment their values: \e{increase_hours} and \e{increase_minutes}. 
Contracts express the semantics:  preconditions (\e{require}), postconditions (\e{ensure}), and class \e{invariant}. 
The postcondition of \e{increase_hours} (line 9) specifies the relation between the values of \e{hours} on routine entry (denoted \e{old hours}) and exit;
the invariant (21 -- 23) states the values' validity ranges. The task of the prover, here AutoProof \cite{tschannen2015AutoProof} --- a prover for contract-equipped program, part of a technology stack that includes the Boogie \cite{le2011boogie} proof engine and the Z3 SMT solver \cite{de2008z3} --- is to establish that every routine execution starting in a state satisfying the precondition and invariant terminates in a state satisfying the postcondition and again the invariant.

\begin{figure}
  \centering
  \includegraphics[width=7cm]{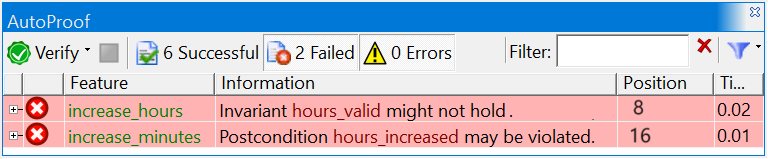}
  \vspace{-0.1cm}
  \caption{AutoProof report of proof failures for class \e{CLOCK}}
  \label{fig: proof_result}
\end{figure}

\begin{figure}[htbp]
\centerline{{\includegraphics[width=7.5cm]{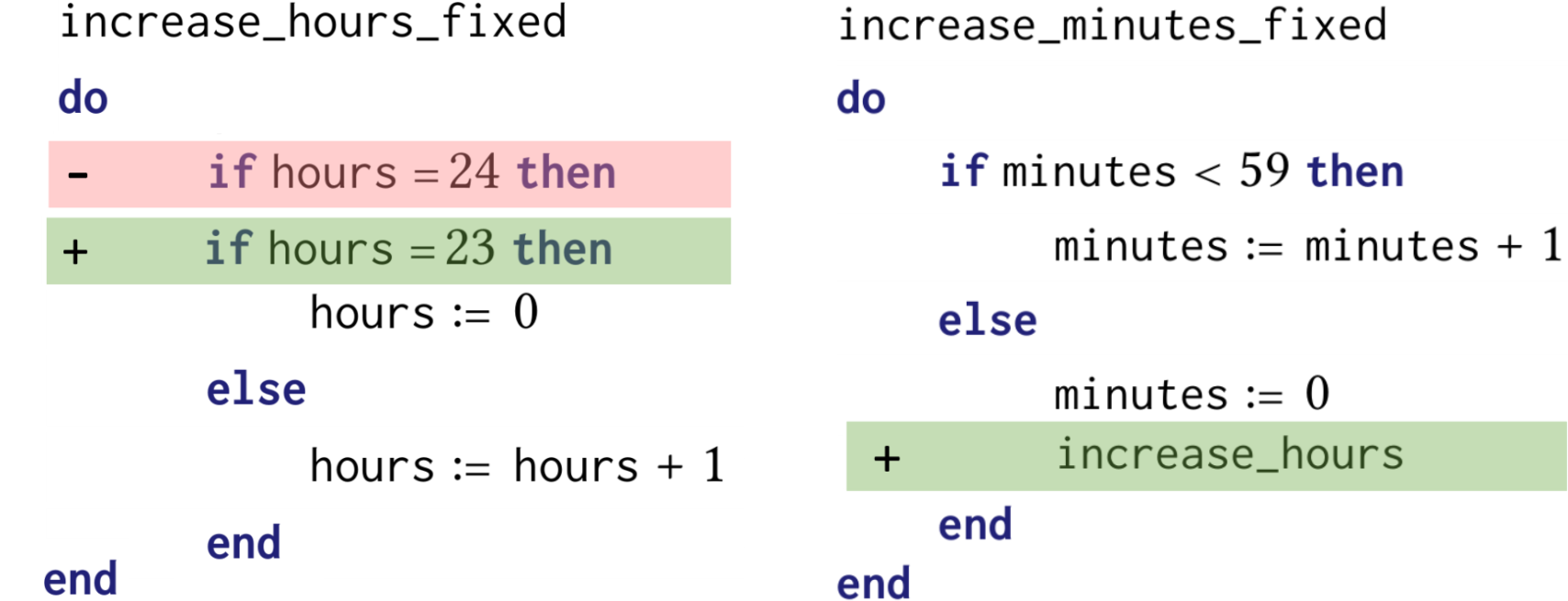}
}}
\vspace{-0.1cm}
\caption{Generated fixes (highlighted)}
\label{fig: clock_fixes}
\end{figure}

The proof attempt with AutoProof fails, producing the messages of Fig. \ref{fig: proof_result}. The first failure results from using 24 instead of 23 as the threshold in line 7; the second one, from forgetting to increase \e{hours} when \e{minutes} reaches 59 (line 14). \toolname~ automatically generates fixes for both bugs, as shown in Fig. \ref{fig: clock_fixes}. The fix for \e{increase_hours} properly identifies the faulty case for \e{hours = 23}, and replaces the  \e{if} condition accordingly; the fix for \e{increase_minutes} correctly calls \e{increase_hours} when \e{minutes = 59}. 
With these two fixes, AutoProof now succeeds in verifying the routines.

\vspace{-0.1cm}

\section{Execution-free fix generation} \label{obtaining}
\begin{figure}[htbp]
\vspace{-0.3cm}
\centerline{{\includegraphics[width=8cm]{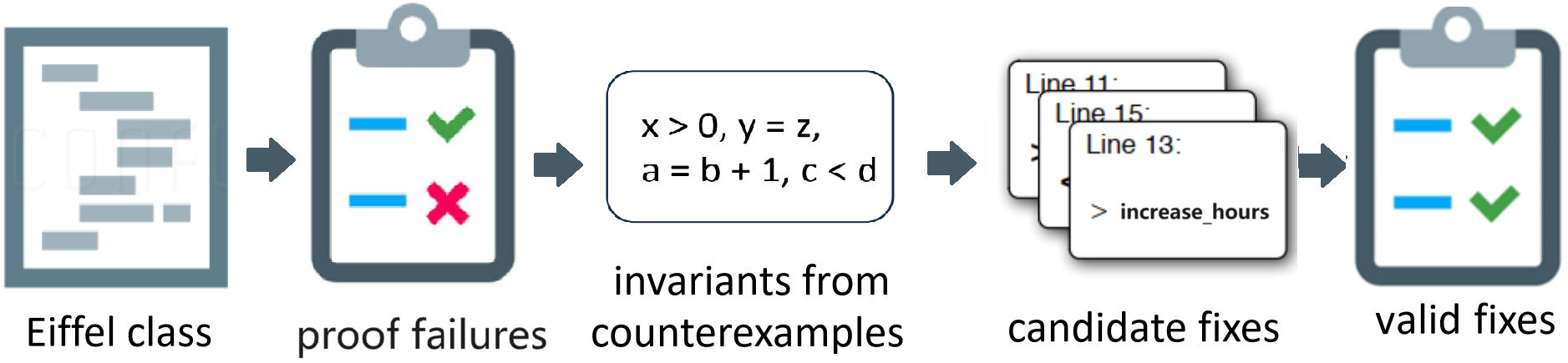}
}}
\vspace{-0.3cm}
\caption{\toolname~ workflow}
\label{fig: workflow}
\end{figure}

Fig. \ref{fig: workflow} summarizes the \toolname~process. AutoProof attempts to verify the class. If it discovers a fault in the program, it reports a proof failure and produces, by leveraging the underlying SMT solver (such as Z3), a set of counterexamples (CEs)\footnote{By running the verification multiple times with different seeds.} illustrating the failure. 
Each contains a trace (a sequence of states) documenting how the program ends up in a state that violates a contract. It then creates \emph{CE invariants}: predicates that always hold in the CEs. 
\toolname~ derives them from \emph{state invariants} ---  predicates that always hold at a certain state in a set of program executions --- which it infers using Daikon \cite{ernst2007daikon}. 

The resulting CE invariants\footnote{"Invariants" as used in Daikon and the present discussion are more general than the OO concept of class invariant (section \ref{example}).} characterize the circumstances for a failure to happen; they guide \toolname's process when generating candidates fixes. 
\toolname~ then runs AutoProof again to retain only the fixes that both remove the original proof failure and introduce no extra failure.
The rest of section \ref{obtaining} details these two key steps: inferring invariants from CEs (\ref{inference}) and synthesizing candidate fixes based on the invariants (\ref{synthesis}).

\vspace{-0.3cm}
\subsection{Inferring CE invariants} \label{inference}

For a class $C$ and one of its routines $r$, let $c_1, ..., c_n$ be the set of CEs produced by the prover when a failure of $r$ occurs.
Each  $c_i$ contains a sequence of states, represented by $s_1^{c_i}, ..., s_m^{c_i}$. Each such state s$_j^{c_i}$ ($1\leq j\leq m$) consists of a set of equalities between program expressions --- including $C$'s attributes (fields) and  $r$'s arguments --- and their values in that state.

To derive invariants that hold in all CEs, \toolname~ feeds the CEs into Daikon, which yields $m$ sets $inv_1, ..., inv_m$, where each $inv_j$ is a set of candidate invariants which hold in the $j^{th}$ state in every CE. Formally:
$$\forall c_i \in \{c_1, ..., c_n\}, s_j^{c_i} \models inv_{j}$$
The derivation relies on a set of predefined templates including:
\begin{itemize}
  \item Equality to primitive-type constants: \e{e = v} where a boolean or integer expression \e{e} has the value \e{v}. In Fig. \ref{fig: clock_fixes}, we get \e{hours = 23} and \e{minutes = 59}.
  \item Non-constant equality: \e{e1} = \e{e2} where \e{e1} and \e{e2} (not both constants) have the same value.
  \item Linear relations: properties of the form \e{e1 = a $\cdot$ e2 + b} or \e{e1 =} \e{a $\cdot$ old e2 + b}, for integer expressions \e{e1}, \e{e2} and constants \e{a}, \e{b}.
\end{itemize}

\noindent The set of invariants obtained from these templates, denoted $P$, is the \emph{basic invariants set}; \toolname~  
derives a \emph{compound invariants set}  $\Pi$ by combining pairs of basic invariants in $P$ through disjunctions. For example, when $P$ = \{\e{hours = 23}, \e{minutes $\geq$ 0}\}, then $\Pi$ = \{\e{hours = 23}, \e{minutes $\geq$ 0}, \e{hours = 23} $\vee$ \e{minutes $\geq$ 0} \}.

The current implementation of Proof2Fix only takes advantages of the invariants in the \emph{initial state}, characterizing a set of faulty inputs that leads to the proof failure and appear to be the most interesting. Future work will explore invariants of other states.

\vspace{-0.3cm}

\subsection{Synthesize candidate fixes}\label{synthesis}
Having obtained CE invariants (elements $\phi$ of the set $\Pi$) 
of a routine $r$, \toolname~ can now produce 
candidate fixes of two kinds, described next: \textit{contract} fixes, which affect the specification of $r$ (contract), and \textit{implementation} fixes, which modify its implementation (body). 

It is generally beyond \toolname's purview to determine which one, of contract and body, should be fixed; if it finds fixes of both kinds, it will let the programmer decide. Empirical studies \cite{raluca} indeed indicate that in practice bugs arise from both kinds of mistake.

\vspace{0.2cm}
\noindent\textbf{Contract fixes} use one of the following strategies:
\begin{itemize}
  \item \textbf{Precondition strengthening}: add ``\e{not} $\phi$'' to $r$'s precondition, to rule out the faulty cases characterized by $\phi$.
  \item \textbf{Postcondition weakening}: let $\psi$ be the postcondition or class invariant that fails in the proof, Proof2Fix replaces it with a weakened version ``\e{not} $\phi$ \e{implies} $\psi$'', making the previously failing cases now verify.
\end{itemize}

\noindent Precondition strengthening requires special care. This strategy is often useful, since forgetting a precondition clause is a common source of bugs (as in forgetting to state that the argument of a real-square-root routine must be non-negative). Applied without restraint, however, it could lead to absurdities; in fact, any routine will verify if we preface it by  \e{require False}. Some cases are more subtle: a proposed fix might inadvertently add a clause \e{not} $\phi$ contradicting  existing clauses (as in a supposed fix that adds the precondition clause \e{a > 0} whereas $r$'s precondition includes \e{a = 0}). 
To filter out such spurious fixes, \toolname~ deliberately injects a fault into $r$ by inserting a contradictory assertion (\e{check False end}\footnote{ {\color[HTML]{222277}\footnotesize\bfseries\ttfamily{check}} {\ttfamily{p}} {\color[HTML]{222277}\footnotesize\bfseries\ttfamily{end}}, equivalent to { \color[HTML]{222277}\footnotesize\bfseries\ttfamily{assert}} {\ttfamily{p}} in some other formalisms, has no effect on execution but verifies only if {\ttfamily{p}} holds at the given program point.}) at the beginning of its body; 
any candidate fix whose verification does not capture the injected fault will be discarded.

\vspace{0.2cm}
\noindent Generating \textbf{Implementation fixes} involves two steps: selecting a fix schema that abstracts common instruction patterns; instantiating the fix schema with proper instructions and $\phi$. 

\begin{figure}[htbp]
\vspace{-0.2cm}
\centerline{{\includegraphics[width=8cm]{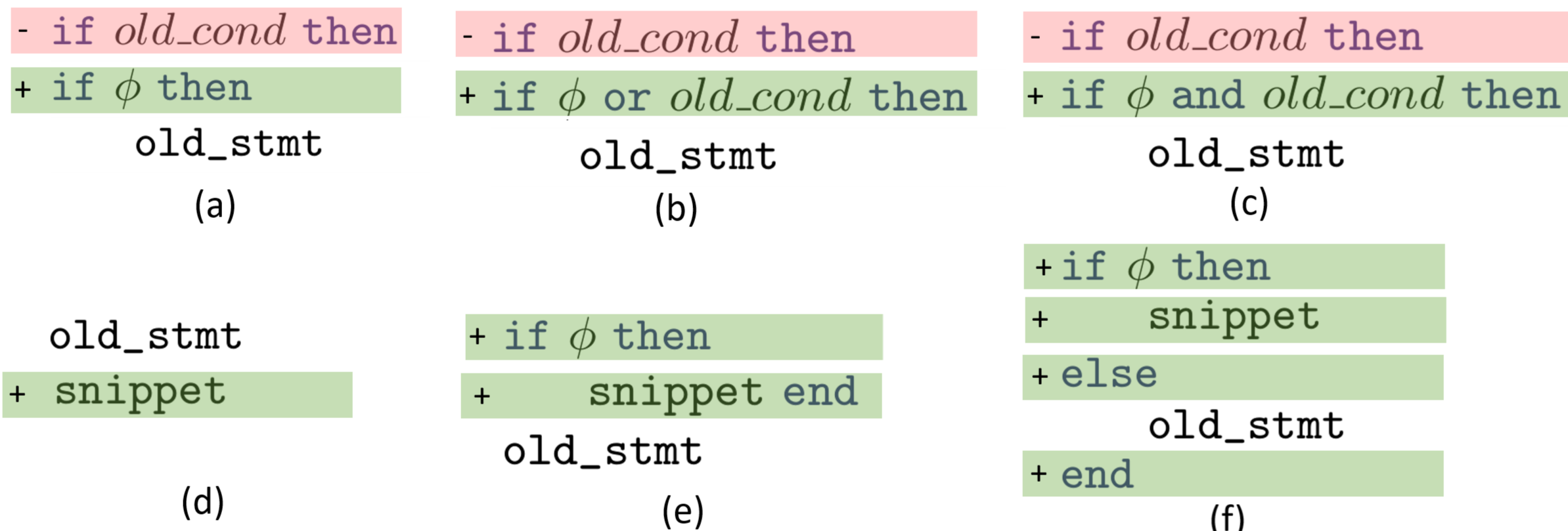}}
}
\vspace{-0.2cm}
\caption{Fix schemas of Proof2Fix}
\label{fig: schema}
\end{figure}

Fig. \ref{fig: schema} shows the currently implemented fix schemas. \e{snippet} is a sequence of instructions, and \e{old\_stmt} represents some instructions in the original program related to the point of failure. \toolname~ assumes that the variables used in the violated postcondition are suspicious; when instantiating \e{snippet}, it selects instructions $i$ that alter the states of the variables, based on their types:
\begin{itemize}
  \item Boolean. $i$ is an assignment \e{e := d} where \e{d} is the constant \e{True} or \e{False} or is the expression \e{not e}.
  \item Integer. $i$ is an assignment \e{e := d} where \e{d} is one of to the constants 0, 1, and $-1$, or the expressions \e{-e}, \e{e + 1}, \e{e - 1}.
  \item Reference. $i$ is a call to a command (procedure) \e{e.p (a$_1$, $...$, a$_n$)}, where \e{p} is a routine available to the faulty routine and \e{a$_1$, $...$, a$_n$} are the arguments.
\end{itemize}
\e{old_stmt} is one of the following:
\begin{itemize}
  \item The single instruction $l$ at the failure's location.
  \item The block of instructions that immediately contains $l$.
\end{itemize}
Accordingly, the instantiated schema replaces the instruction at the failure's location or the whole block. The two fixes in Fig. \ref{fig: clock_fixes} are generated by instantiating the fix schemas (a) and (d).

\noindent To improve performance when $r$ contains multiple blocks, \toolname~ identifies those involved in all CEs and treats them in priority, as the ones most likely to be faulty.


\section {Evaluation} \label {evaluation}
This section reports a preliminary assessment of the quality of the automatically generated fixes from \toolname~ on a collection of faulty Eiffel programs\footnote{All code and results are available at \url{https://github.com/proof2fix/proof2fix}.}, including data-structure classes adapted from an old version of EiffelBase\footnote{https://www.eiffel.org/doc/solutions/EiffelBase}, examples in the AutoProof tutorial\footnote{http://AutoProof.sit.org/AutoProof/tutorial}, benchmarks of previous software verification competitions \cite{bormer2011cost,klebanov20111st}, and a benchmark of recursive programs\footnote{https://github.com/maple-repair/recursive-benchmark}. 
The experiment ran on a Windows 11 machine with a 2.10 GHz Intel 12-Core processor and 32 GB of RAM. The number of CEs used in invariant inference for each failure is set to 10 empirically. 
The EiffelBase examples (started with ``\e{V\_}'' in Table \ref{table_fixing_result}) are particularly significant as they are real bugs (not seeded), which actually arose in earlier released versions of the library, and were subsequently corrected\footnote{Since the technology has evolved, some of the examples would no longer compile in their original form; we re-injected the original bugs into the current versions.}. 
\vspace{-0.3cm}

\begin{table}[htbp]
 \scriptsize
  \centering
   \renewcommand\arraystretch{1.1}
  \caption{Fixing Results of Proof2Fix}
   \vspace{-0.3cm}
    \begin{tabular}{|c|ccccc|}
    \hline
    Classes & LOC & \#Failures & \#Valid & 
    \#Proper & Avg.T$_f$ (s) \\
    \hline
    ABSOLUTE & 18 & 2 & 2 & 2 & 49
    \\ \hline
    CONSEQ & 17 & 4 & 4 & 3 & 55
    \\ \hline
    INCREMENT & 17 & 4 & 4 & 2 & 49
    \\ \hline
    MAX & 17 & 4 & 4 & 1 & 71
    \\ \hline
    MIN & 26 & 4 & 4 & 1 & 64
    \\ \hline
    SUM & 20 & 8 & 4 & 2 & 48
    \\ \hline
    ACCOUNT & 102 & 7 & 5 & 2 & 54
    \\ \hline
    CLOCK & 141 & 9 & 9 & 5 & 89
    \\ \hline
    ARITHMETIC & 190 & 4 & 3 & 2 & 50
    \\ \hline
    HEATER & 78 & 4 & 3 & 3 & 169
    \\ \hline
    LAMP & 81 & 4 & 4 & 0 & 147
    \\ \hline
    J\_ABSOLUTE & 41 & 12 & 12 & 5 & 51
    \\ \hline
    MAX\_IN\_ARRAY & 40 & 5 & 0 & 0 & 58
    \\ \hline
    SQUARE\_ROOT & 45 & 4 & 3 & 0 & 47
    \\ \hline
    V\_ARRAY & 67 & 1 & 1 & 1 & 61
    \\ \hline
    V\_ARRAYED\_LIST & 76 & 1 & 0 & 0 & 91
    \\ \hline
    V\_INDEXABLE\_CURSOR & 123 & 1 & 1 & 0 & 98
    \\ \hline
    V\_LINKED\_LIST & 38 & 2 & 2 & 1 &  78
     \\ \hline
    \textbf{Total} & 1137 & 80 & 66 & 30  & 70
    \\ \hline
    \end{tabular}%
  \label{table_fixing_result}%
\end{table}%
\vspace{-0.3cm}

\noindent Table \ref{table_fixing_result} lists, for each class, its length in lines of code (LOC), the number of failures (\#Failures) detected by AutoProof, and for how many of those failures \toolname~ built (at least one) valid (\#Valid) or proper (\#Proper) fixes, as well as the average fix generation time (in seconds) for each failure (Avg.\#T$_f$). 

The difference between the last two categories comes from the usefulness (in the current state of the approach) of a human reasonableness check. A fix is valid if the repaired program passes verification. Some valid fixes, however, might be over-enthusiastically change the intended semantics of the program; the most important case is the weakening of a postcondition, which is sometimes justified but sometimes results in a useless program. For that reason, we currently perform a manual check of valid fixes and discard any that would distort the intuitively understood intent of the code. \textit{Proper} fixes are valid fixes that pass this human check. Valid but improper fixes include: too-restrictive precondition strengthening; too-generous postcondition strengthening (as already mentioned), including an implication ``\e{not} $\phi$ $\implies$ $\psi$'' where $\phi$ in fact always evaluates to \e{true}; a spurious fix that recursively calls the routine itself to ensure its postcondition. Note that improper valid fixes remain valuable, not as fixes but as debugging aids, since they clearly evidence failure-inducing inputs.

\begin{figure}[htbp]
\vspace{-0.2cm}
\centerline{{\includegraphics[width=5cm]{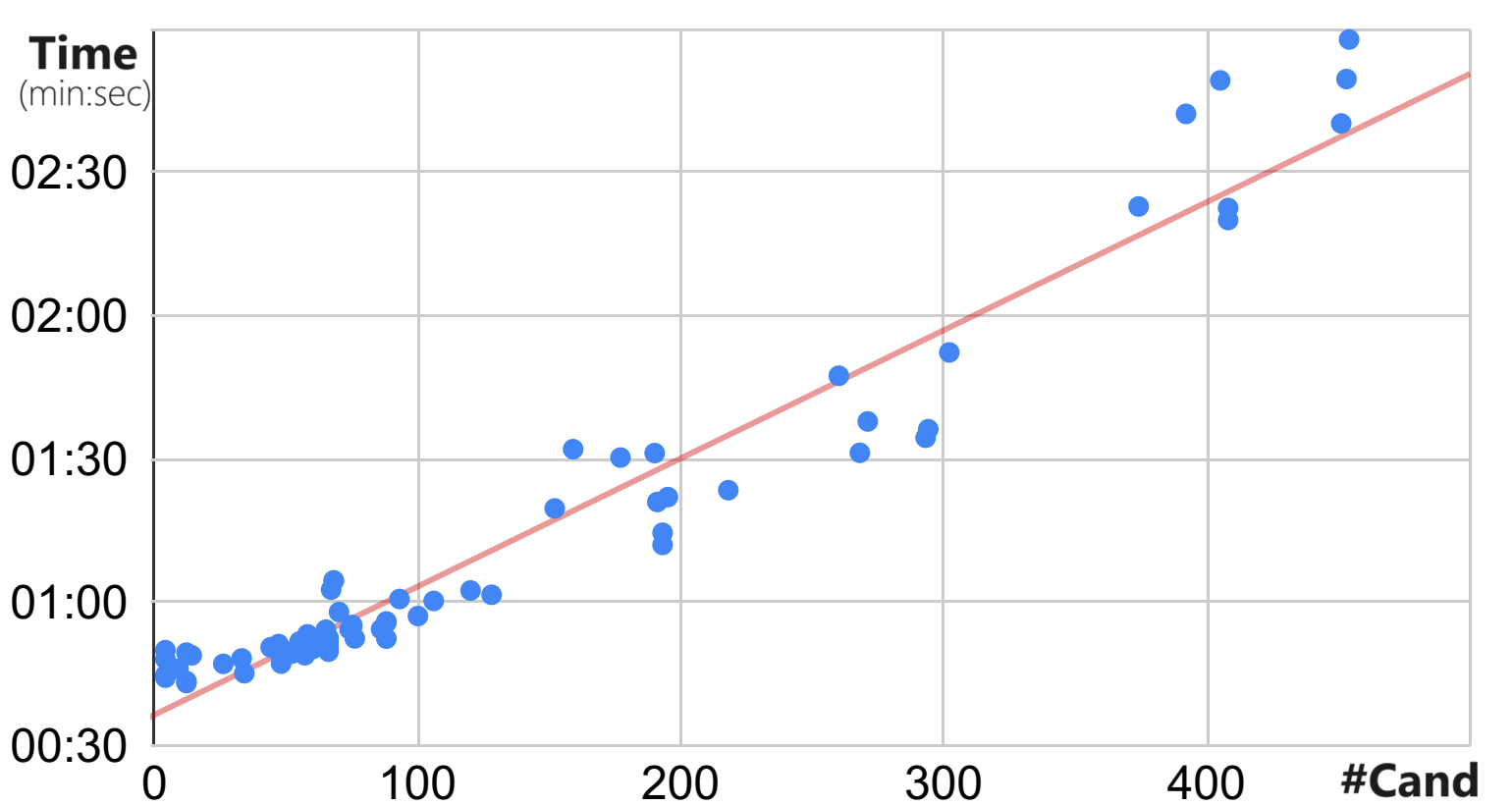}
}}
\vspace{-0.2cm}
\caption{Fix time over number \#Cand of candidate fixes 
}
\label{fig: time_vs_cand}
\end{figure}

Among the 80 failures detected by AutoProof, \toolname~ is able to generate at least one valid fix for 82.5\% of them and at least one proper fix for 37.5\% of the failures.
On average, \toolname~ ran for 70 seconds for each failure, which seems an acceptable overhead. 
For each of the 66 fixed failures, \toolname~ generated, on average, 124 candidates and 5 valid fixes; the average percentage of valid fixes per candidate is 4\%. 
Fig.\ref{fig: time_vs_cand} shows the correlation between the fix time and the number of candidate fixes (\#Cand); the increment of \#Cand would not lead to significant increase of time: when \#Cand $<$ 100, time increases very slowly; when \#Cand is beyond 100, time roughly grows linearly.

\noindent Table \ref{table: failure_types} correlates the causes of proof failures with the effectiveness of \toolname~ in finding valid fixes for that category of failures. 

\vspace{-0.3cm}

\begin{table}[htbp]
 \scriptsize
  \centering
   \renewcommand\arraystretch{1.1}
  \caption{Failure causes and fixes}
  \vspace{-0.3cm}
    \begin{tabular}{|p{120pt}|ccc|}
    \hline
    Cause of Failure  & \#Failure & \#Fixed & \#Proper Fixed \\
    \hline
    incorrect expression of an assignment & 25 & 21 (84\%) & 12 (48\%)
    \\ \hline
    incorrect condition of an if statement & 18 & 17 (94\%) & 12 (67\%)
    \\ \hline
    weakness of assisting contracts (postcondition of a callee or loop invariant) & 10 & 3 (30\%) & 0 (0\%)
    \\ \hline
    missing instruction (assignment or routine call) & 7 & 7 (100\%) & 3 (43\%)
    \\ \hline
    missing precondition & 6 & 5 (83\%) & 2 (33\%)
    \\ \hline
    incorrect postcondition & 4 & 4 (100\%) & 0 (0\%)
    \\ \hline
    incorrect algorithm & 7 &  6 (100\%) & 1 (14\%)
    \\ \hline
    others & 3 & 3 (100\%) & 0 (0\%)
    \\ \hline
    \end{tabular}%
  \label{table: failure_types}%
\end{table}%
\vspace{-0.3cm}

\noindent From these results, \toolname~ is most effective for failures caused by incorrect expression of an assignment and incorrect condition of an \e{if} statement. In both cases, \toolname~ can obtain CE invariants that accurately characterize the faulty cases that need to be ruled out; \toolname~ is not good at fixing failures caused by weakness of assisting contracts, as the CEs generated in those cases usually vary a lot, making it difficult to infer useful invariants.

\section{Previous work} \label{previous}
AutoFix~\cite{6776507, 7203042} performs automated fixing of programs based on patterns for fixing the program.
The main differences with \toolname~ are that it locates faults by using dynamic program analysis (which is more time consuming) instead of proofs, and it proposes fixes based on rather simple heuristics and verifies them using a test suite (instead of proving the resulting code).
A tool based on similar ideas is JAID~\cite{chen2017contract}, which first extracts the contracts, then applies fixing and validation techniques similar to those of AutoFix.
Other tools such as SemFix~\cite{nguyen2013semfix} or Nopol~\cite{7463060} also use test suites as a verification tool for fixes generated through symbolic execution.
Nilizadeh \cite{9438573} showed that this is prone to overfitting. 
In contrast, \toolname's fixes are guaranteed to be correct, a result that no test-based approach can provide. In addition, the prover-based approach provides precise location identification for the failure.  

Closer to \toolname~are FootPatch~\cite{10.1145/3180155.3180250},  Maple~\cite{nguyen2019automatic}, and the approach of Logozzo and Ball~\cite{10.1145/2398857.2384626}, which use a formal approach for identifying faults, creating and verifying fixes.
These three tools address and fix specific classes of faults: memory errors for (FootPatch); issues that can be translated to linear arithmetic expressions (Maple); contracts, initialization, method purity and guards (Logozzo and Ball).
\toolname~ is more generic and has the potential to fix any fault discovered by the prover and verify the fixes.




\section{Planned developments} \label{future_plans}

Our goal is to make \toolname~ part of a standard development environment such as EiffelStudio (the main Eiffel IDE).
Although the experiment results have demonstrated the effectiveness of the Proof2Fix in fixing bugs on various examples, including (most convincingly in our view) historical bugs in production code, several further developments are necessary for the approach to deliver its full value.

\noindent {\bf Optimization of CE invariants.}
The quality of generated fixes from Proof2Fix relies on the quality of CE invariants. We are working on more sophisticated kinds of invariants, including first-order predicates with quantifiers. We are also improving the predicate sets by spotting and removing redundancies and contradictions.





\noindent {\bf Fix schema generality.} The current implementation does not instantiate fix schemas with more than one predicate at a time and empirically limits the maximum admitted length of a snippet to a small number. We are developing more sophisticated techniques.

\noindent {\bf Bug and fix diversity.} More generally, a major goal of our current extension work is to expand the range of bug and fix types dramatically, by performing an extensive review of historical bugs and corrections in large software repositories and deriving a set of patterns covering as many of them as possible, based on the conjecture that while the set of real-world bugs and fixes is large, many of them can be attached to a tractable number of categories.

\noindent {\bf Fix ranking.} \toolname~ often finds several valid fixes for a given failure. We are working towards ranking them properly.

\noindent {\bf Number of CEs}
The maximum number of CEs is, so far, fixed. We are devising rigorous, empirically validated criteria for choosing the best value for every case.

\noindent {\bf IDE integration} With the promise of tools such as \toolname, Automatic Program Repair can become, rather than an esoteric facility to be run on demand, a component of the normal practice of developing, verifying and debugging programs. We are working towards including \toolname~ as an integral part of a production-grade software development environment.



\section{Conclusions} \label{conclusions}

Programmers deserve Automatic Program Repair: mechanisms to suggest correct, trustworthy and reliable fixes to the mistakes they inevitably made, and to mistakes inevitably made by any AI-based tools that become part of their tool base. 
Using a completely static approach, based on a powerful state-of-the-art program-proving tool stack, overcomes the inherent limitations of dynamic (test-based approaches): it is faster, does not require preparing test cases, works on simulated environments (particularly for embedded and cyberphysical systems), and --- perhaps even more importantly ---  guarantees that proposed fixes are valid corrections. 
We believe that \toolname, as presented in this article, provides a promising step towards making Automatic Program Repair a realistic everyday component of modern software development. 

\bibliographystyle{ACM-Reference-Format}
\bibliography{reference}

\end{document}